\documentclass[useAMS,usenatbib]{mn2e}
\usepackage[cp1251]{inputenc} %
\usepackage{graphicx}
\usepackage{amssymb}
\usepackage[english]{babel}
\usepackage{morefloats}
\usepackage[colorlinks,citecolor=blue]{hyperref}
\usepackage{ulem}
\usepackage[dvipsnames]{color}

\def\ltsima{$\; \buildrel < \over \sim \;$}
\def\simlt{\lower.5ex\hbox{\ltsima}}
\def\gtsima{$\; \buildrel > \over \sim \;$}
\def\simgt{\lower.5ex\hbox{\gtsima}}

\newcommand{\ssfr}{\ensuremath{\Sigma_{\rm SFR}\,}}
\newcommand{\ssfrcl}{\ensuremath{\Sigma_{\rm SFR,\, cl}\,}}
\newcommand{\ssfruv}{\ensuremath{\Sigma_{\rm SFR,\, UV}\,}}
\newcommand{\sgas}{\ensuremath{\Sigma_{\rm gas}}\,}

\newcommand{\Xcounit}{cm\ensuremath{^{-2} } (K km s\ensuremath{^{-1} })\ensuremath{^{-1} }\,}
\newcommand{\kmps}{km~s\ensuremath{^{-1}}\,}
\newcommand{\Msun}{M\ensuremath{_\odot}\,}
\newcommand{\Msunpc}{M\ensuremath{_\odot}~pc\ensuremath{^{-2}}\,}

\newcommand{\Msunyrpc}{M\ensuremath{_\odot}~yr\ensuremath{^{-1}}~pc\ensuremath{^{-2}}\,}
\newcommand{\Oo}{\displaystyle}

\title[A Kennicutt-Schmidt relation]{A Kennicutt-Schmidt relation at molecular cloud scales and beyond}

\author[Khoperskov \& Vasiliev]{Sergey A. Khoperskov$^{1,2}$\thanks{sergey.khoperskov@obspm.fr} \& Evgenii O. Vasiliev$^{3,4}$ \\
$^1$ GEPI, Observatoire de Paris, PSL Research University, CNRS, Universit\'{e} Paris Diderot, Sorbonne Paris Cit\'{e}, \\ Place Jules Janssen, 92195 Meudon, France \\
$^2$Institute of Astronomy, Russian Academy of Sciences, Pyatnitskaya st., 48, 119017 Moscow, Russia \\
$^3$Southern Federal University, Sorge 5, Rostov on Don 344090, Russia \\
$^4$Special Astrophysical Observatory, Russian Academy of Sciences, Nizhnii Arkhyz, Karachaevo-Cherkesskaya Republic,  369167 Russia
}

\begin{document}

\label{firstpage}

\maketitle
\begin{abstract}
Using $N$-body/gasdynamic simulations of a Milky Way-like galaxy we analyse a Kennicutt-Schmidt relation, $\ssfr\propto \sgas^N$, at different spatial scales. We simulate synthetic observations in CO lines and UV band. We adopt the star formation rate (SFR) defined in two ways: based on free fall collapse of a molecular cloud  -- \ssfrcl, and  calculated by using a UV flux calibration -- \ssfruv.  We study a KS relation for spatially smoothed maps with effective spatial resolution from molecular cloud scales to several hundred parsecs. We find that for spatially and kinematically resolved molecular clouds the $\ssfrcl \propto \sgas^N$ relation follows the power-law with index $N \approx 1.4$.  Using UV flux as SFR calibrator we confirm a systematic offset between the \ssfruv and \sgas distributions on scales compared to molecular cloud sizes. Degrading resolution of our simulated maps for surface densities of gas and star formation rates we establish that there is no relation $\ssfruv - \sgas$ below the resolution $\sim 50$~pc. We find a transition range around scales $\sim 50-120$~pc, where the power-law index $N$ increases from 0 to $1-1.8$ and saturates for scales larger $\sim 120$~pc. A value of the index saturated depends on a surface gas density threshold and it becomes steeper for higher $\sgas$ threshold. Averaging over scales with size of $\simgt 150$~pc the power-law index $N$ equals $1.3-1.4$ for surface gas density threshold $\sim 5$~\Msunpc.  At scales $\simgt 120$~pc surface SFR densities determined by using CO data and UV flux, $\ssfruv/\ssfrcl$, demonstrate a discrepancy about a factor of $3$. We argue that this may be originated from overestimating (constant) values of conversion factor, star formation efficiency or UV calibration used in our analysis.  
\end{abstract}


\section{Introduction}\label{sec::intro}

Established empirically a Kennicutt-Schmidt~(KS) relation quantitatively tell us how much molecular gas is required to support star formation at a given rate in a galaxy \citep{1959ApJ...129..243S,1998ApJ...498..541K}.  Further observational studies demonstrate that this relation remains valid for galaxies of various type and mass \citep{2008AJ....136.2846B,2008AJ....136.2782L,2013AJ....146...19L}.  The growth of multiwavelength data for different galaxies and the increase of spatial resolution lead to a transition from studying disk-averaged star formation to sub-kpc scales \citep[see also for review,][]{2007ARA&A..45..565M,2012ARA&A..50..531K}. 

Using the CO and H$\alpha$ datasets \citet{2010ApJ...722L.127O} found in M~33 that the star formation and gas surface densities correlate well at $1$~kpc resolution, meanwhile the correlation becomes  weaker with higher spatial resolution and it breaks down at giant molecular cloud~(GMC) scales. In the recent study \citet{2015A&A...578A...8B} came to the same conclusions based on more data in different wavelengths including FUV as well. Analysing the spatial distribution of $\rm CO/H_\alpha$ peaks in M~33 \cite{2010ApJ...722.1699S} established that the scaling relation between gas and star formation rate surface density observed at large scales does not have its direct origin in an instantaneous cloud-scale relation. This consequently produces a breakdown in the star formation law as a function of the surface density of the starforming regions.  Obviously, the relation is believed to be violated at small scales due to the drifting young clusters from their parental GMCs~\citep{2010ApJ...722L.127O} or mechanical stellar feedback effects~\citep{2013ApJ...779....8K} or the role of turbulence \citep{2014ApJ...784..112K} or other mechanisms.  From another point of view, \cite{2012ApJ...758..127F} found that at low molecular gas surface density and on sub-kpc scales, an accurate determination of the slope on the basis of CO observations will be difficult due to uncertainties of $\rm CO/H_2$ conversion factor. Thus, it is claimed that a KS relation is valid only on scales larger than that of GMCs, when the spatial offset between GMCs and star forming regions is smoothed \citep{2010ApJ...722L.127O}, and the relation holds only for averaging over sufficiently large scales \citep[e.g.,][]{2012ARA&A..50..531K,2014ApJ...786...56B}.

Observational evidences for the spatial offset between molecular gas and star forming regions have been found for M~51 \citep{2011ApJ...726...85E,2013ApJ...779...42S,2013ApJ...763...94L} and four nearby low-luminosity AGNs \citep{2015A&A...577A.135C}. Deviations from the KS-type relation have been found on small scales~($\leq 100-200$~pc) at low gas densities~\citep{2010ApJ...722.1699S}. Following by \citet{2007ApJ...671..333K}, who demonstrated a variation of power-law index, \citet{2015A&A...577A.135C}  has also found that a KS relation can be either sublinear or superlinear: the slopes range from 0.5 to 1.3 and increase for larger spatial scale. The observed spatial scale at which a KS relation has a breakdown is ranged from $100$ to $500$~pc~\citep[e.g.,][and others]{2010ApJ...722L.127O}. Such scatter may be originated due to both physical and systematic effects, one of them that multiwavelength datasets used in the analysis are frequently obtained for different spatial resolution, e.g. it is varied from $50$ to $200$~pc~\citep[e.g.,][]{2016A&A...587A..44P}. In general, a variation of the break down scale can be considered by using a kind of uncertainty principle \citep{2014MNRAS.439.3239K}, which states that a break down of starformation--density relation on small spatial scales is expected to be due to the  incomplete sampling of both starforming regions and initial mass function and the spatial drift between gas and stars. 

One part of a KS relation, gas surface density \sgas, is determined by atomic and molecular hydrogen densities. The former is obtained in HI surveys, the latter is not directly observed and it can be found from CO measurements assuming some CO-H$_2$ conversion factor \citep[e.g.,][]{1998ApJ...498..541K}. The other part of the relation, surface SFR density \ssfr, can be defined in several ways. At first, a \ssfr value can be found by estimating free-fall time for a given collapsing cloud. This approach is usually used in numerical simulations, where both mass and volume gas density can be determined. Observationally a surface SFR density value is obtained by converting various SFR calibrations/estimators in photometric bands and spectral lines, e.g., FUV, IR, FIR, H$\alpha$, P$\alpha$ and etc~\citep[see for review,][]{2012ARA&A..50..531K}. These calibrations may be used alone or combined in composite tracers. Some of them are connected not only with stellar population, but also with recombinations in ionized gas. Young stars emit enormous UV photons, so that UV flux is a direct SFR estimator. Certainly, observations of UV radiation are accompanied by many physical obstacles mainly connected to interstellar dust attenuation, but in numerical simulations one can transfer radiation correctly to get UV radiation field and, as a consequence, surface SFR density.

Here based on our numerical simulations of galactic evolution~\citep{2016MNRAS.455.1782K} we analyse how a KS relation behaviours on sub-galactic scales and aim to find a spatial scale at which the relation has a breakdown. We investigate in detail how the relation depends on spatial resolution of our synthetic observations. The paper is organized as follows. Section~\ref{sec::model} describes our model in brief. Section~\ref{sec::results} present our results. Section~\ref{sec::concl} summarizes our findings.

\section{Model}\label{sec::model}

To simulate the galaxy evolution we use our code based on the unsplit TVD MUSCL (Total Variation Diminishing Multi Upstream Scheme for Conservation Laws) scheme for gas dynamics and the $N$-body method for stellar component dynamics. Stellar dynamics is calculated  by using the second order flip-flop integrator. For gravity calculation we use the TreeCode approach. A complete description of our code can be found in \citet{2014JPhCS.510a2011K,2016MNRAS.455.1782K}. Below we describe it in brief with a particular stress on chemical kinetics and radiation transfer parts.

We implemented the self-consistent cooling/heating processes~\citep{2013MNRAS.428.2311K} coupled with the chemical evolution of 20 species including CO and H$_2$ molecules using simplified chemical network described by~\citet{1999ApJ...524..923N}. Based on our simple model for H$_2$ chemical kinetics~\citep{2013MNRAS.428.2311K} we expand the \citet{1999ApJ...524..923N} network by several reactions needed for hydrogen ionization and recombination. For H$_2$ and CO photodissociation we use the approach described by~\citet{1996ApJ...468..269D}. The CO photodissociation cross section is taken from \citet{2009A&A...503..323V}. In our radiation transfer calculation described below we get ionizing flux at the surface of a computational cell. To calculate self-shielding factors for CO and H$_2$ photodissociation rates and dust absorption factor for a given cell we use local number densities of gas and molecules, e.g. $f_{sh}^{\rm H_2} = n_{\rm H_2} L$, where $n_{\rm H_2}$ is H$_2$ number density in a given cell and $L$ is its physical size. The chemical network equations is solved by the CVODE package~\citep{Hindmarsh2005}. We assume that a gas has solar metallicity with the abundances given in \citet{2005ASPC..336...25A}: $[{\rm C/H}] = 2.45 \times 10^{-4}, [{\rm O/H}] = 4.57\times 10^{-4}, [{\rm Si/H}] = 3.24\times 10^{-5}$. Dust depletion factors are equal to $0.72$, $0.46$ and $0.2$ for C, O and Si, respectively. We suppose that silicon is singly ionized and oxygen stays neutral. 

For cooling and heating processes we extend our previous model~\citep{2013MNRAS.428.2311K} by CO and OH cooling rates \citep{1979ApJS...41..555H} and CI fine structure cooling rate~\citep{1989ApJ...342..306H}. The other cooling and heating rates are presented in detail in Table~2~\cite[Appendix B in][]{2013MNRAS.428.2311K}. Here we simply provide a list of it: cooling due to recombination and collisional excitation and free-free emission of hydrogen~\citep{1992ApJS...78..341C}, molecular hydrogen cooling \citep{1998A&A...335..403G}, cooling in the fine structure and metastable transitions of carbon, oxygen and silicon \citep{1989ApJ...342..306H}, energy transfer in collisions with the dust particles \citep{2003ApJ...587..278W} and recombination cooling on the dust \citep{1994ApJ...427..822B}, photoelectric heating on the dust particles \citep{1994ApJ...427..822B,2003ApJ...587..278W}, heating due to H$_2$ formation on the dust{ particles}, and the H$_2$ photodissociation \citep{1979ApJS...41..555H} and the ionization heating by cosmic rays \citep{1978ApJ...222..881G}. In our simulations we achieve gas temperature value as low as 10~K and number density as high as $5\times 10^3$~cm$^{-3}$. 

In the star formation recipe adopted in our model mass, energy and momentum from the gaseous cells, where a star formation criteria are satisfied~(local Jeans instability, converging flow, temperature threshold), are transited directly to newborn stellar particles. To compute mass loss, energy feedback and UV emission radiated by stellar population we use the stellar evolution code STARBURST'99~\citep{1999ApJS..123....3L} assuming solar metallicity of stars and Salpeter IMF with mass limits of 0.1 and 100~\Msun. 

We render the UV radiation from young stars by tracing the ultraviolet photon rays on the fly. To account molecule photodestruction we should know spatial structure of UV background in the galactic disc. Recent observations provide some evidences for significant radial and azimuthal variations of UV flux in the nearby galaxies~\citep{2007ApJS..173..185G}. No doubt that such variations are stipulated by local star formation. So that we need to include radiation feedback from stellar particles in our calculations.

Through our simulations the UV emission of each stellar particle is computed with the stellar evolution code {\sc STARBURST'99}~\citep{sb99} assuming solar metallicity of stellar population. So that for each particle we know its luminosity evolution. After that we separate particles in two groups: young stellar particles (the age is smaller than $20$~Myr) and the other ones.
For definiteness we assume a uniform background field ten times lower than that in the Solar neighbourhood, $F_{b}=0.1$~Habing. Thus the UV background $F^{\rm UV}$ in a hydrodynamical cell with coordinates $\Oo {\bf {\rm r}_0}$ can be written as
\begin{equation}
\label{eq::UVequation}
 F_{\rm UV}({\bf {\rm r}_0}) = F_{\rm b} +  \sum_{i} F^{\rm old}_{i}({\rm r}_0) + \sum_{j} F^{\rm young}_{j}({\rm r}_0,{\rm r}_j)\,,
\end{equation}
where $\sum_{i} F^{\rm
old}_{i}({\rm r}_0)$ is deposit from old stellar population (age $>20$~Myr), which plays a role only locally, in a cell where the stellar particle locates~($\Oo {\bf {\rm r}_0}$). The last term is UV flux from young stellar population -- the brightest stars. Their deposit is the most important in photodestruction of molecules in surrounding medium.

Due to small number of young stellar particles at each integration time step, we can use the ray-tracing approach for each stellar particle. For $j$-th "young particle" we estimate the{ radius of} spherical shell (similar to the Stroemgren sphere), where the UV field value decreases down to $0.1$~Habing:
\begin{equation}
\Oo R^{\rm d}_j = 0.1 \delta \sqrt{L^*_j/ (4\pi)}\,,
\end{equation}
where $L^*_j$ is luminosity of $j$-th stellar particle in Habing units and $\delta $ is the effective cell size. For each shell we calculate the UV flux assuming the optical depth $\tau = 2  N / (10^{21} \rm{cm^{-2}})$,
where $N$ is the total column density of gas in cm$^{-2}$. So that we can obtain the distribution of the UV intensity in the entire galactic disc according to Eq.~\ref{eq::UVequation}.

Previously we simulated the evolution of galaxies having different morphological type \citep{2016MNRAS.455.1782K}. Here we constrain our study by one model, because of similar results for the others. We consider the model of a disk galaxy with a bar and four spiral arms --- model B in our list of models in \citet{2016MNRAS.455.1782K}, which mimics the Milky Way morphology.  This model is characterized by the constant star formation rate $4.5-5$~\Msun~yr$^{-1}$ during the first $600$~Myr of evolution. Similar to our previous study here we analyse the galaxy at $500$~Myr. At this moment we have $2\times10^6$ stellar particles of different age. 

The spatial resolution in our simulations of gas dynamics is 4~pc, which is reasonably smaller than that can be reached in the observations aimed to study the KS relation \citep[e.g.,][]{2010ApJ...722L.127O,2015A&A...577A.135C,2015A&A...578A...8B}. Such high numerical resolution allow us to separate molecular cloud structure and star forming clusters and follow the relation on scales from individual clouds and clusters to kpc-sizes. 

\section{Results}\label{sec::results}

\subsection{A KS relation based on molecular cloud free-fall time}

To make synthetic observations we compute the CO line emission maps with post processing radiation transfer approach. The physical parameters of molecular clouds are extracted by applying CLUMPFIND method for simulated CO line spectra~\citep{1994ApJ...428..693W}. Here we adopt brightness temperature threshold $T^{\rm th}_{\rm b} = 1$~K and spectra resolution is $\delta v = 0.1$~\kmps. Thus, we have physical parameters (size, mass, total CO luminosity and velocity dispersion) for each cloud in the sample of 813 clouds~\citep[see details in][]{2016MNRAS.455.1782K}. 

We follow the standard notations from~\cite{2012ARA&A..50..531K} for star formation rate and gas surface density.
A molecular gas surface density in a pixel with coordinates ${\bf r}$ is calculated as follow:
\begin{equation}
\label{eq::gas_density}
\sgas({\bf r}) = W_{\rm CO}({\bf r}) \times X_{\rm CO}\,\,{\rm [M_\odot\, pc^{-2}]} \,,
\end{equation}
where $W_{\rm CO}$ is the CO line integrated intensity, $X_{\rm CO}$ is the $\rm CO-H_2$ conversion factor, which is adopted for simplicity to be constant and  equal to $2\times10^{20}$~\Xcounit~\citep[e.g.,][]{2013ARA&A..51..207B}, although many studies give evidences of its variation in different environments~\cite[see for review][]{2012ARA&A..50..531K}. Note that such simplification is commonly used both in interpretation of observation data and in simulations. Moreover, here we analyse the Milky Way size galaxy assuming a constant metallicity $Z=Z_\odot$. Because the conversion factor value adopted is determined for the Milky Way gas, our simplification is believed to be reasonable enough.

A global star formation rate for a given collapsing cloud can be estimated as~\citep{2005ApJ...630..250K}:
\begin{equation}
 \label{eq::sfr_individual_clouds}
 \Oo \ssfr = \varepsilon M_{\rm cl}t^{-1}_{\rm ff}\,\, {\rm [M_\odot\, yr^{-1}]}\,,
\end{equation}
where $\varepsilon=0.014$ is the star formation efficiency, $t_{\rm ff}$ is the free-fall time. The rate \ssfrcl cannot be compared to the gas surface density $\sgas({\bf r})$ directly, because the former is determined for the whole cloud. Then, we smooth the \ssfrcl value over cloud surface by taking into account brightness distribution within the cloud:
\begin{equation}
 \label{eq::sfr_clouds}
 \Oo \ssfrcl({\bf r}) = \ssfr\frac{W_{\rm CO}({\bf r})}{L_{\rm CO}}\,\, {\rm [M_\odot\, yr^{-1} kpc^{-2}]}\,,
\end{equation}
where $L_{\rm CO}$ is the total cloud luminosity in CO lines. Obviously, an integrated SFR over the cloud surface is equal to \ssfr. 

\begin{figure}
\centering
\includegraphics[width=1.0\columnwidth]{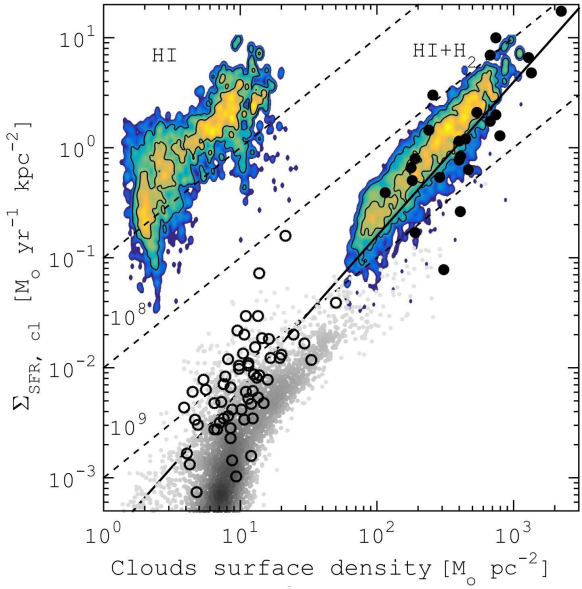}
\caption{ The relation between gas surface density and star formation rate found for identified clouds. The right contour map shows the relation based on the total (HI+H$_2$) gas surface density, whereas the left ones corresponds to the dependence for HI gas only. The color reflects the number of clouds, the contours correspond to the 30, 60 and 90\% from the maximum value in the distribution. The grey small points correspond to the observational data obtained by \citet{2010AJ....140.1194B}, the large circles are taken form original paper by \citet{1998ApJ...498..541K}, where open and filled ones correspond to normal and starburst galaxies, respectively. The power law with slope $N=1.4$ is shown by solid line. Dashed lines correspond to constant gas depletion time $10^7$, $10^8$, and $10^9$~yr.}
\label{fig::sfr_hih2}
\end{figure}

Figure~\ref{fig::sfr_hih2} presents the relation between gas surface density, \sgas, and star formation rate,  \ssfrcl, found for the molecular clouds derived in our simulation. One can clearly see that the dependence for our simulated data follows the power law with slope $N=1.4$ (compare to solid line) and the locus of the observational points used by \citet{1998ApJ...498..541K} for establishing his relation coincides with our contour map based on the total ($\rm HI+H_2$) gas surface density. 

\subsection{A KS relation based on UV calibration}
Using the ray-tracing technique we obtained UV brightness maps  in our simulations on the fly \citep{2016MNRAS.455.1782K}. Since UV flux is a direct tracer of young stellar population, star formation rate is usually estimated by using the well-known calibration~\citep{1998ApJ...498..541K}:
\begin{equation}
\label{eq::sfr_uv}
\rm \Sigma_{\rm SFR, UV}({\bf r}) = 1.4\times 10^{-28} L_{\rm UV}\,\, {\rm [M_\odot\, yr^{-1} kpc^{-2}]}
\end{equation}
where the coefficient is calculated using Salpeter IMF with mass limits of $0.1$ and $100$~\Msun, which in turn is in agreement with our star formation and feedback prescriptions, {\bf $\rm L_{\rm UV} = F_{\rm UV} / S$ is the UV luminosity and $\rm S$ is the pixel surface, the flux is taken from the Eq.~\ref{eq::UVequation}.}

\begin{figure*}
\centering
\includegraphics[width=0.95\hsize]{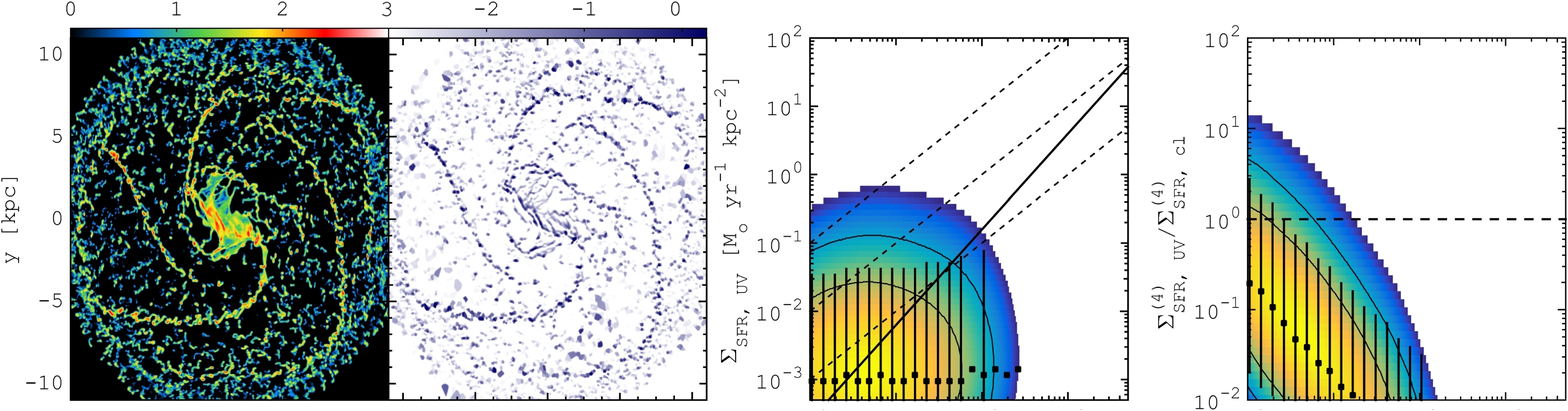}\vskip-0.005\vsize
\includegraphics[width=0.95\hsize]{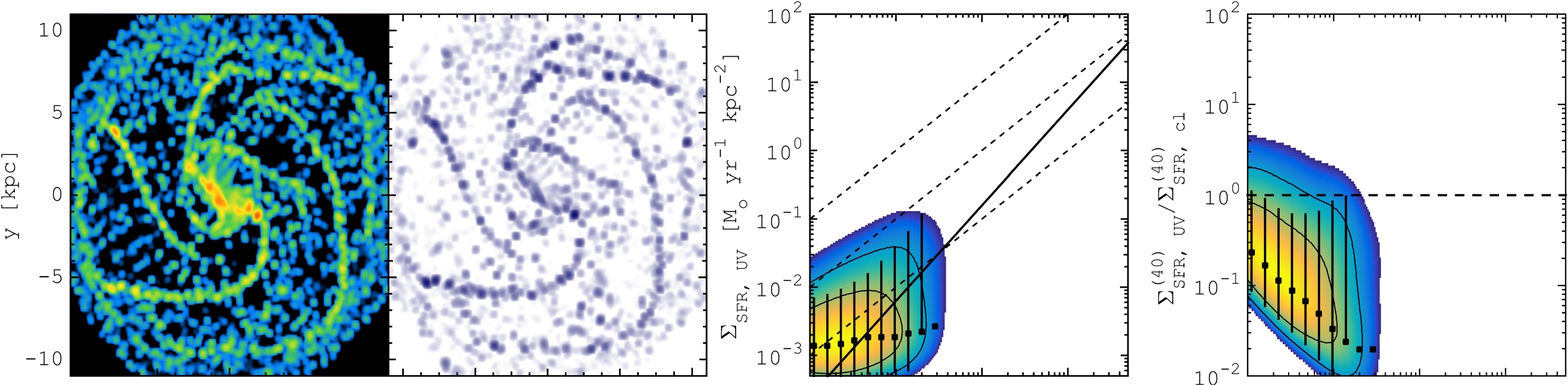}\vskip-0.005\vsize
\includegraphics[width=0.95\hsize]{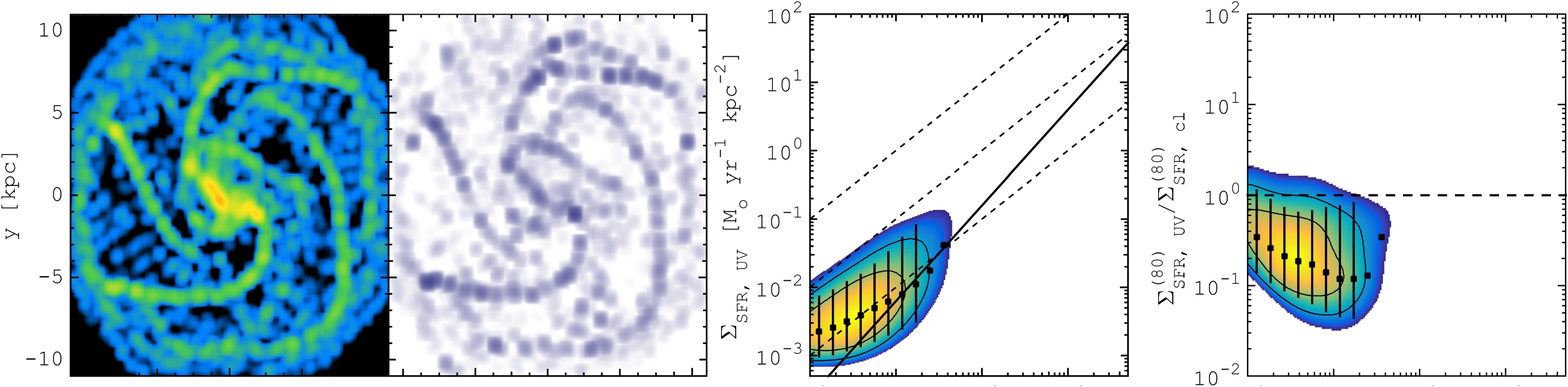}\vskip-0.005\vsize
\includegraphics[width=0.95\hsize]{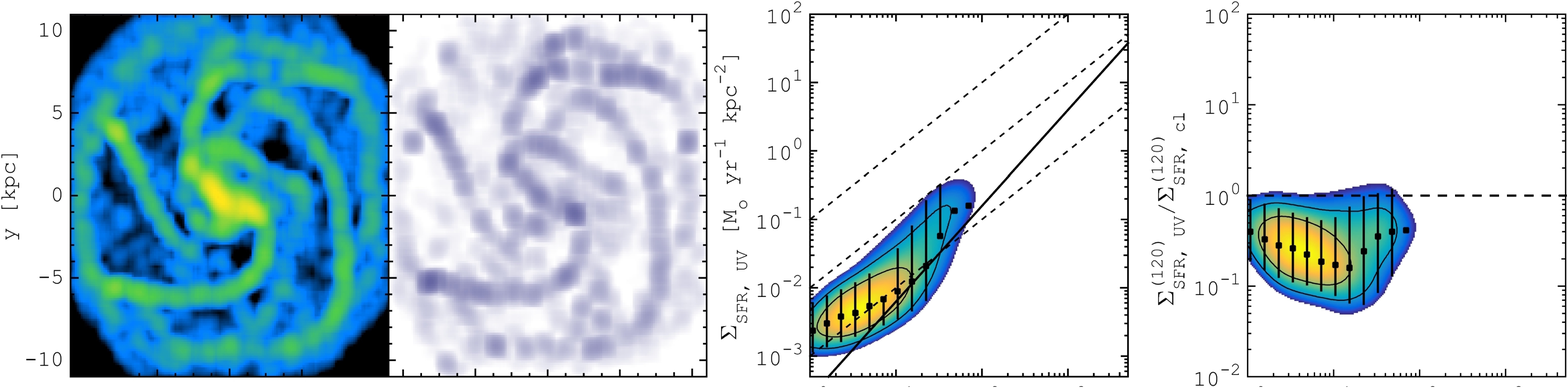}\vskip-0.005\vsize
\includegraphics[width=0.95\hsize]{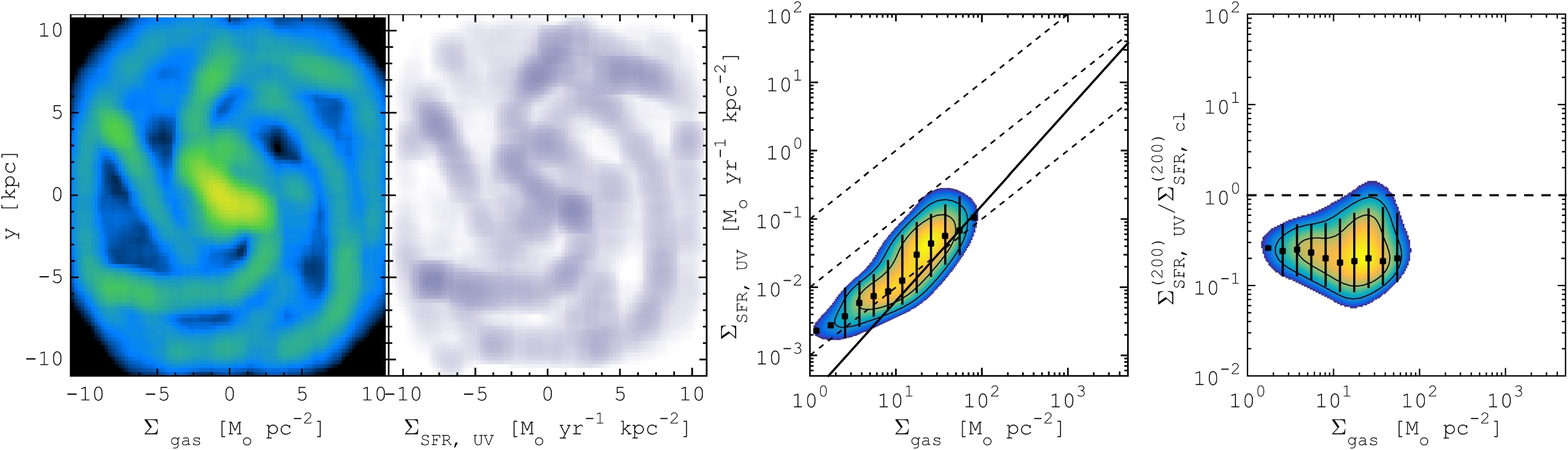}
\caption{
The gas surface density $\Sigma^{(n)}_{\rm gas}({\bf r})$ (left column of panels) and star formation rate estimated by UV flux $\Sigma^{(n)}_{\rm SFR, UV}({\bf r})$ (left middle column of panels)  maps for convolution kernel $n= 4, 40, 80, 120$ and $200$~pc, respectively, from top to bottom panels. The distributions for $n=4$~pc correspond to the original data. The $\ssfruv - \sgas$ relation (third column of panels). The dashed lines show constant gas depletion time $10^7, 10^8, 10^9$~yrs from top to bottom, respectively, the solid line corresponds to the original KS relation with index $N=1.4$~\citep{1998ApJ...498..541K}.  The dependence of the $\ssfruv/\ssfrcl$ ratio on $\sgas$  is depicted on right column of panels. The black squares in both right columns depict the mean value of the distributions, the error bars show 70\% scatter.}
\label{fig::convolution}
\end{figure*}

Top row of panels in Figure~\ref{fig::convolution} presents gas surface density and UV flux snapshots of the galaxy (two left panels). The gas surface density and UV emission maps follow the large-scale structure of the galaxy and look very similar. However, there is a systematic offset between them on small scales. That is easily confirmed by the absence of correlation between \ssfruv and \sgas~(right middle panel). The surface SFR remains almost constant at level $\sim 10^{-3}$~\Msunyrpc and shows a significant scatter around it. That means there are no UV sources in a gas with the CO brightness temperature threshold $T^{\rm th}_{\rm b} \geq 1$~K. By our definition such regions are considered as a part of molecular clouds. 
Note also that due to a negligible UV flux in the inner parts of giant molecular clouds there are no points for \sgas higher than $\sim 50$~\Msunpc in the $\ssfruv \propto \sgas$ dependence presented in the third column in Figure~\ref{fig::convolution}, whereas this range is filled in the $\ssfrcl \propto \sgas$ relation depicted in Figure~\ref{fig::sfr_hih2}. 

Certainly, young bright UV sources can photodissociate molecular gas up to several tens parsecs around \citep[the exact size depends on both physical conditions in a gas, and UV flux, see details][]{1996ApJ...458..222B}, so any correlation can be hardly expected in our pixel-by-pixel analysis. That confirmed by a strong anticorrelation between ratio \ssfruv/\ssfrcl \ and \sgas \ presented in the right panel (top row of Figure~\ref{fig::convolution}).  Obviously, if we expand in size a region, where CO and UV luminosity values are compared, then we find a characteristic size, which embraces both molecular gas and young stellar particles. That corresponds to observations with relatively low spatial resolution, then, we should average (convolve) our gas surface density and UV emission maps.

\subsection{A KS relation averaged over sub-galactic scales}

In recent extragalactic CO and UV emission observations structures smaller than ten parsecs are hardly to be resolved. Then here we consider the beam smearing effect by reducing the quality of our simulated maps, in other words, by degrading spatial resolution. At first, we convolute the distributions of gas surface density (\sgas, Eq.~\ref{eq::gas_density}), star formation rates based on  free fall time collapsing model (\ssfrcl, Eq.~\ref{eq::sfr_clouds}) and star formation rate based on  UV emission (\ssfruv, Eq.~\ref{eq::sfr_uv}) as following:
\begin{equation}
 \label{eq::sfr_clouds_conv}
 \Oo Y^{(n)}({\bf r}) = \int Y({\bf r'}) H({\bf r - r'}, n) d{\bf r'} \,\,
\end{equation}
where function $Y({\bf r})$ is $\sgas({\bf r})$, $\ssfruv({\bf r})$, or $\ssfrcl({\bf r})$; $H({\bf r},{\bf r'}, n)$ represents the Gaussian kernel with a half-width, which is equal to $n$ parsecs. Note that the distributions for $n=4$~pc correspond to the original images. At second, we re-bin the smoothed maps with a bin size equal to $n$ pc.

Figure~\ref{fig::convolution} presents the convoluted maps of gas surface density and UV surface SFR (two left columns), the $\ssfruv \propto \sgas$ relation and the $\ssfruv/\ssfrcl$ ratio vs. gas density $\sgas$~(two right columns) for kernel size $n=4, 40, 80, 120, 200$~pc (from top to bottom rows). For convolution kernel up to $n=40$~pc (second row in Fig.~\ref{fig::convolution}) there is no significant correlation between $\ssfruv$ and $\sgas$. Of course, one can note an appearance of small positive slope in the $\ssfruv \propto \sgas$ relation and a decrease of the scatter around the average value. This reflects that several stellar clusters and molecular clouds are separated by a distance around $40$~pc. Such groups are expected to be young and located in the densest environment of galactic disk ($\sgas \simgt 50$~\Msunpc).  The increase of the kernel size to $n=80$ and even $120$~pc leads to a remarkable $\ssfruv \propto \sgas$ relation, its slope depends on $\sgas$ value: it is closer to $1-1.4$ for higher surface density and the scatter decreases (third and fourth rows). So that one can conclude that a scale around $100$~pc is a critical and a KS relation is expected to be appeared beyond this spatial scale value. This scale is sufficiently large so that a majority of molecular clouds have nearby stellar particle counterpart independently on environmental density. For a typical drift velocity value $\sim 10-20$~\kmps a stellar cluster needs about $5-10$~Myr to overcome the distance around $100$~pc. This timescale is short enough for remaining stellar particle young and saving neighbouring molecular clouds against destruction by the photodissociating radiation of  young stellar particle. For largest kernel size considered here, $n=200$~pc, the relation is well-defined with almost constant slope around unity. 

The degrading resolution procedure leads to smoothing both surface gas density and star formation rate based on UV emission distributions. This has at least two consequences. The first is that the spatial offset between gas surface density and UV emission distribution becomes smaller and it may disappear at all for larger convolution kernel size. Obviously, this occurs when the kernel size is comparable or larger than a distance between molecular clouds and neighbouring starforming regions and their sizes as well. So that increasing kernel size we lost the information on small scales related to local starforming processes and come to global ones -- those that we usually detect in real observations. The second is that in some number of cells there are surface gas density values below the threshold used for the cloud definition. Taking into account both consequences we can analyse how the degrading resolution influences on the $\ssfruv \propto \sgas$ dependence. 

\begin{figure}
\centering
\includegraphics[width=1\hsize]{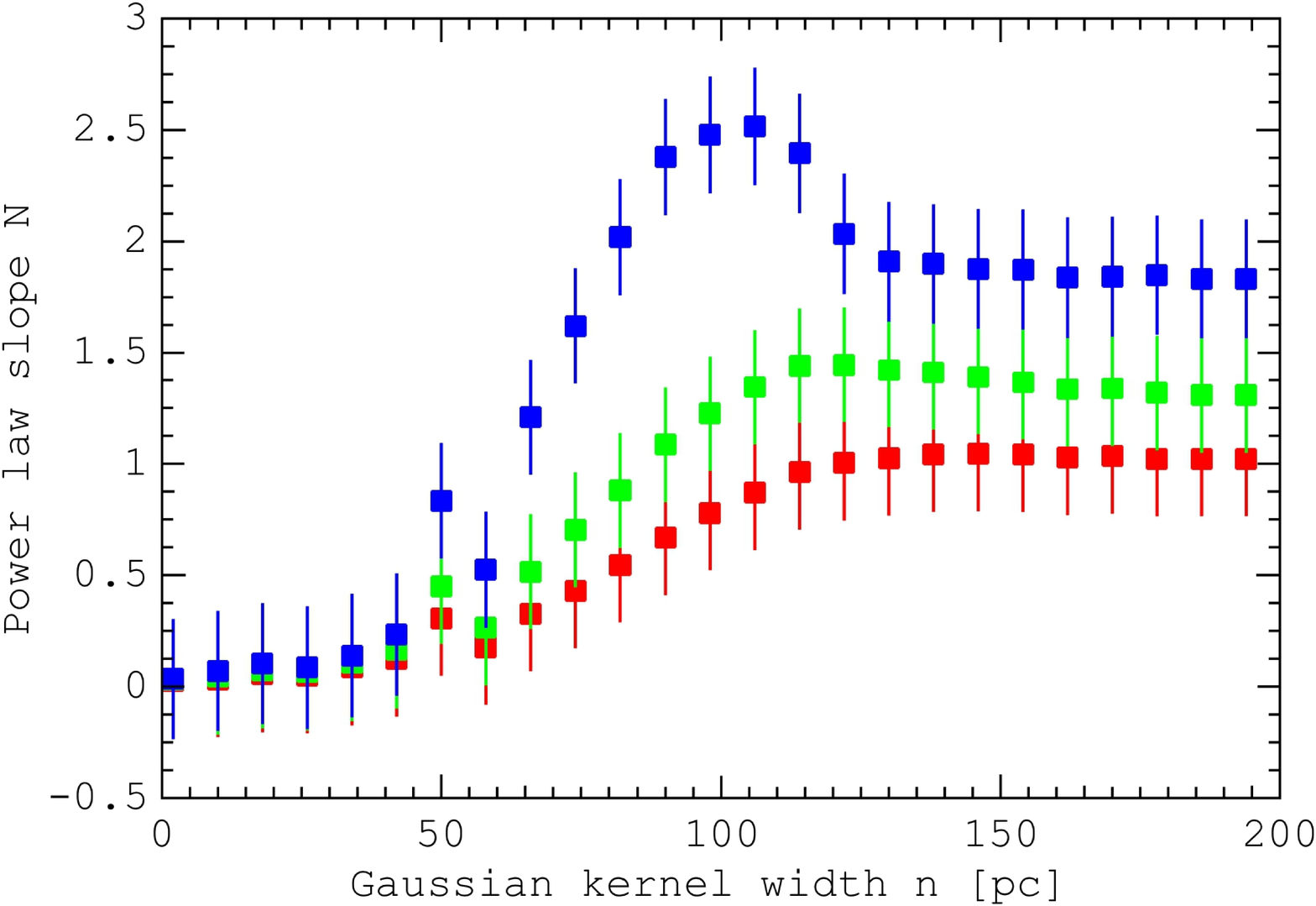}
\caption{
The dependence of the power-law slope $N$ in the $\ssfruv \propto \sgas^N$ relation on convolution kernel size $n$ for different surface density thresholds: $0.1$~\Msunpc (red), $5$~\Msunpc (green) and $10$~\Msunpc (blue).  The symbols depict the mean value of the distributions, the error bars correspond to $70$\% scatter.
}
\label{fig::slope}
\end{figure}

Figure~\ref{fig::slope} presents more detailed dependence of the slope in the $\ssfruv \propto \sgas^N$ relation on convolution kernel size $n$ pc. Choosing high surface gas density threshold we consider cells corresponded to the densest parts of molecular clouds, after the convolution procedure we 're-distribute' the surface values (gas density and starformation rate) over neighbouring cells. In case of extremely high threshold the offset between the surface values remains significant and total number of cells involved in our analysis is small, then an error in estimating $\ssfruv - \sgas$ slope appears to be large. On the other hand for quite low surface gas density threshold we can take into account atomic gas unrelated to molecular clouds \citep[e.g.,][]{2016MNRAS.455.1782K}. So that here we consider intermediate values for gas surface density threshold.

In Figure~\ref{fig::sfr_hih2} one can note that the gas surface  density in molecular clouds is usually higher than $\sim 5$~\Msunpc (see the lowest contours on the right color map). Then, we can consider a dependence on surface density threshold, here we adopt three values equal to $0.1$, $5$ and $10$~\Msunpc. For any level one can find that the $\ssfruv - \sgas$ has no dependence below the resolution $\sim 50$~pc, a transition range around $\sim 50-120$~pc, where the slope increases from zero up to $1-1.8$, and the saturation of the index in this range for resolution larger as $\sim 120$~pc. Note that for kernel size $\sim 50-120$~pc the power-law index becomes harder for higher \sgas \ threshold and reaches $2.5$ for the highest threshold value at $\sim 100$~pc. Such behaviour can be explained that when we constrain surface density threshold, then denser clouds are taken into consideration or, in other words, the number of clouds decreases (see contours on the map in Figure~\ref{fig::convolution}). For larger kernel the index demonstrates a small bump, it is clearly seen for the highest $\sgas$  threshold considered here. Going to larger scales molecular clouds from neighbour starforming regions are included to our analysis, then the slope decreases slightly. One can conclude that a mean distance between evolutionary independent starforming regions is around $120$~pc. The power-law index is around $1.3-1.4$ for  the gas surface density threshold higher than $\sim 5$~\Msunpc. This slope is very close to the index in the original KS relation obtained for averaging over  the whole galaxy. Note that the data depicted in Figure~\ref{fig::convolution} corresponds to the smallest threshold, so that the slope obtained for that data is around unity.

Note that we investigated the dependence of the $\ssfruv - \sgas$  slope for gas surface density varied up to $100$~\Msunpc and found that the mean value for the $\ssfruv - \sgas$ slope does not exceed 2, but the error in estimating the $\ssfruv - \sgas$ slope becomes high enough for $\simgt 20$~\Msunpc and reaches more than $\pm 1$ around $1.5-2$, whereas it is constrained by $\pm 0.25$ around $1-1.8$ for $\sgas \sim 0.1-10$~\Msunpc (Figure~\ref{fig::slope}). That is a result of small number of cells included into the analysis (see the description of the consequences of the degrading resolution procedure above).

Using our simulations a SFR value can be found by two different ways: one is based on estimating free-fall time value (Eq.~\ref{eq::sfr_clouds}) and the another is adopted by using UV calibration. Right column of Figure~\ref{fig::convolution} presents how the $\ssfruv/\ssfrcl$ ratio depends on \sgas. For the original resolution one can see simple anticorrelation, because the typical distance between molecular clouds and UV sources in our simulations is larger than $4$~pc. The increase of kernel size larger than $120$~pc leads to that the ratio becomes more flatten. For instance, it is almost constant for the size of $\simgt 150$~pc. However one can note that the ratio is systematically below unity. That means that the $\ssfruv$~(or $\ssfrcl$) is under- (over-)estimated about a factor of $3$~(for the kernel size $200$~pc). Because of no dependence on $\sgas$ this discrepancy may be originated from using some incorrect constant factor for estimating $\sgas$, $\ssfrcl$ and $\ssfruv$, e.g., that may be conversion factor value in Eq.~\ref{eq::gas_density} or star formation efficiency $\epsilon$ in Eq.~\ref{eq::sfr_individual_clouds}.  Note that different UV calibrations may be also considered for estimating star formation rate in Eq.~\ref{eq::sfr_uv}.

\section{Conclusions}\label{sec::concl}

Based on the 3D simulations of the galactic evolution we analyse how the relation between surface star formation rate (\ssfr) and surface gas density (\sgas) -- a Kennicutt-Schmidt relation -- depends on spatial scale. We study a KS relation for a Milky Way-like galaxy and follow the dependence from inner structure of molecular clouds to several hundred parsecs. We analyse synthetic observations in both CO line and UV band with different spatial resolution. To determine \ssfr we consider two different ways: one is based on estimating free-fall time for molecular cloud collapse -- $\ssfrcl$, and the other is found by using the well-known UV calibration  \citep{1998ApJ...498..541K} -- $\ssfruv$. Our results can be summarized as follows.
\begin{enumerate}
 \item The $\ssfrcl \propto \sgas^N$ relation obtained by using the simulated CO line emission maps follows the power law with index $N=1.4$, the locus of the simulated relation coincides with the observational points used by \citet{1998ApJ...498..541K} for establishing his relation. 
 \item Using UV flux as SFR calibrator one can find a systematic offset between the \ssfruv and \sgas distributions on scales compared to molecular cloud sizes. Averaging over different spatial scales we find (a) there is no dependence $\ssfruv - \sgas$ below $\sim 50$~pc; (b) a transition range around $\sim 50-120$~pc, where the power-law index in the relation increases  from $0$ to $1-1.8$; (c) there is a saturation of the index for spatial resolution larger than $\sim 120$~pc. 
\item For spatial resolution $\sim 50-120$~pc the power-law index becomes steeper for higher $\sgas$ threshold. One can conclude that a mean distance between evolutionary independent star forming regions is around $120$~pc. The power-law index is around $1.3-1.4$ for surface gas density threshold higher than $\sim 5$~\Msunpc, which is typical for molecular clouds. 
\item The ratio of surface SFR densities determined by two different ways, $\ssfruv/\ssfrcl$, is flatten becomes constant in the range $1-100$~\Msunpc at spatial scales $\simgt 120$~pc. However, it is three times lower than unity. This discrepancy may be explained by varying conversion factor $X_{\rm CO}$~(see Eq.~\ref{eq::gas_density}) and/or star formation efficiency $\epsilon$~(see Eq.~\ref{eq::sfr_individual_clouds}) and/or UV calibration~(see Eq.~\ref{eq::sfr_uv}).
\end{enumerate}

\section{Acknowledgments}
We wish to thank the referee for thoughtful suggestions that have improved the quality of the paper. The numerical simulations have been performed at the Research Computing Center~(Moscow State University) under the Russian Science Foundation grant~(14-22-00041). This work was supported by the Russian Foundation for Basic Research grants (14-02-00604, 15-02-06204, 15-32-21062). SAK thanks the ANR (Agence Nationale de la Recherche) MOD4Gaia project (ANR-15-CE31-0007) and the President of RF grant (MK-4536.2015.2). EOV is thankful to the Ministry of Education and Science of the Russian Federation (project 3.858.2017) and RFBR (project 15-02-08293). The thermo-chemical part was developed under support by the Russian Science Foundation (grant 14-50-00043).

\bibliography{mycloudbib_sf1}

\end{document}